# Lasing from a circular Bragg nanocavity with an ultra-small modal volume

Jacob Scheuer,[a)] William M. J. Green,[b)] Guy A. DeRose,[b)] Amnon Yariv [b)]

*Abstract—* We demonstrate single-mode lasing at telecommunication wavelengths from a circular nanocavity employing a radial Bragg reflector. Ultra-small modal volume and Sub milliwatt pump threshold level are observed for lasers with InGaAsP quantum well active membrane. The electromagnetic field is shown to be tightly confined within the 300nm central pillar of the cavity. The quality factors of the resonator modal fields are estimated to be on the order of a few thousands.

The quest for the ultimate localization of light has been one of the central directions in contemporary research in many fields such as integrated optics, quantum communication and computation, sensing and more.[1-3] Optical nanocavities with high quality factors ($Q$) and small modal volume ($V_{mode}$) are key elements for a wide variety of applications such as functional building blocks for integrated optical circuits, lasers, optical traps, and optical logic.[4, 5]

In the past few years, much attention was focused on decreasing the modal volume and improving the $Q$ of photonic crystal (PC) defect cavities by carefully optimizing the position, dimensions and shape of the holes composing the crystal.[6-9] This optimization procedure attempts to tune the effective length of the cavity to the maximal reflection frequency of the PC reflector and is generally conducted numerically. The main disadvantage of this process is the enormous number of parameters to be optimized and the endless number of configurations that potentially need to be considered.

Recently, we have proposed and demonstrated a novel class of circular resonators that are based on optimally designed radial Bragg reflectors.[9-11] These devices, known as annular Bragg resonators (ABRs), are designed to support azimuthally propagating modes, with energy concentrated within a radial defect region or in a central pillar by radial Bragg reflection. Compared to conventional resonators based on total-internal-reflection (TIR), the employment of the Bragg reflection mechanism offers improved control over the resonator parameters ($Q$, $V_{mode}$, etc.) and allows for engineering unique mode profiles. Compared to PC defect resonators, the radial symmetry of ABRs allows for *analytical* engineering the Bragg reflector for optimized $Q$ and smallest modal volume. These properties make the ABR structure highly suitable for realizing ultra-compact cavities, especially if they are designed for the mode with angular modal number $m$=0. The $m$=0 mode is an interesting solution of a disk Bragg resonator consisting of a disk surrounded by a radial Bragg stack (see Fig. 1). Unlike other solutions with nonzero angular modal numbers, the $m$=0 mode is non-degenerate and features maximum intensity at the center of the device. This mode cannot be supported in conventional cavities because the propagation direction of the waves is perpendicular to the cavity interfaces.

In this paper we report on the observation of single mode lasing from a Brag-based disk nanocavity with ultra-small modal volume, realized in active semiconductor material. Figure 1 depicts a scanning electron microscope (SEM) image of the nanocavity which was designed to resonate in the $m$=0 angular mode. The internal disk can be viewed as the cavity while the concentric rings form the radial Bragg reflector, both designed to efficiently confine the specific mode in the cavity. The radius of the optimized nanocavity for telecommunication wavelengths is approximately 150nm.

For a circular cavity to support a radial mode with an angular propagation coefficient of $m$, the radius of the inner disk, $\rho_0$, and the grating profile $\Delta\varepsilon$ must satisfy[9]:

$$\Delta\varepsilon(\rho) = \begin{cases} -2|\alpha| & \sin\{2\text{phase}(H_m^{(1)}[k_0 n_{eff}\rho])\} < 0 \\ 0 & \sin\{2\text{phase}(H_m^{(1)}[k_0 n_{eff}\rho])\} \geq 0 \end{cases}, \quad (1)$$

$$J_m(k_0 n_{eff} \rho_0) = 0$$

where $k_0$, $n_{eff}$ and $\alpha$ are respectively the desired resonance wavenumber, the slab effective index and the perturbation strength. $H_m$ and $J_m$ are respectively the Hankel and the Bessel functions of order $m$. The widths of the resulting layers are determined by the zeros and extrema of the Bessel function of order $m$. Due to the cylindrical geometry, the optimal layer widths required to confine the light in the central pillar are not constant[9, 11, 12] but rather monotonically decreasing with the radial distance. The chirped nature of the gratings, which can be clearly seen in Fig. 1, is highly important. Deviation from that profile will result in non-optimal phase relations between the partially reflected waves from the Bragg layers, and thus, in weaker confinement of the electromagnetic field. It should

[a)]Center for the Physics of Information, California Institute of Technology, 1200 East California Boulevard, Mail Code 128-95, Pasadena, California 91125; koby@caltech.edu
[b)]Department of Electrical Engineering, California Institute of Technology, 1200 East California Boulevard, Pasadena, California 91125

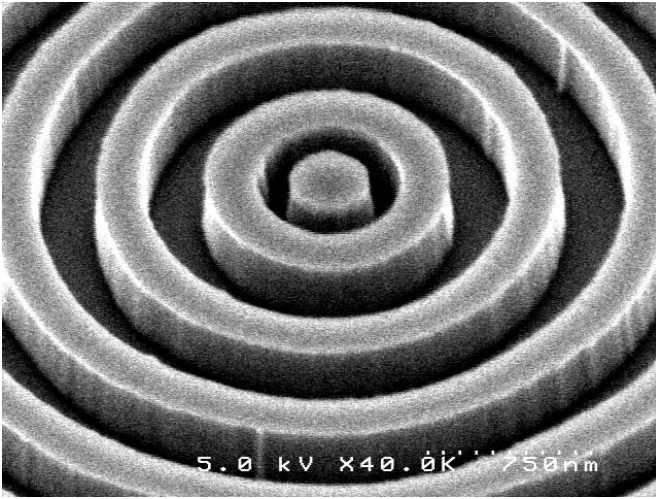

Fig. 1. SEM image of the circular Bragg nanocavity designed to support the $m=0$ mode in the 300nm wide central pillar.

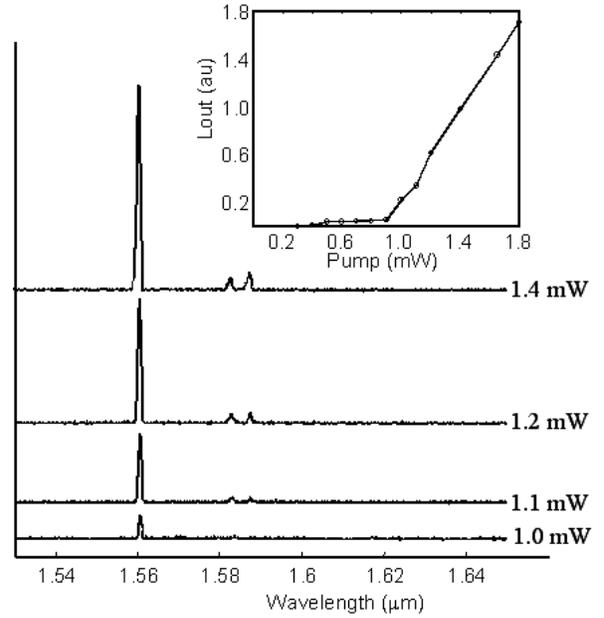

Fig. 2. The evolution of the emitted spectrum from the device shown in Fig. 1 as a function of the pump intensity. Inset – L-L curve, indicating a lasing threshold of $P_{th}=900\mu W$.

be noted that the realized grating profile shown in Fig. 1 is of "second Bragg order" in the sense that the second term in the Fourier expansion of the grating is the term responsible for the radial reflection. Such an approach induces efficient vertical emission from the device which decreases the overall $Q$ on one hand but enables simple observation of the intensity pattern evolving in the device on the other. The condition on the radius of the pillar can be intuitively understood when the optimized distributed reflector is considered effectively as a perfect mirror at the resonant Bragg $k$-vector[8]. The modal field profile of the device is given by a Bessel function of order $m$ in the central pillar and an exponentially decaying Bessel function in the grating area where the decay constant, $k$, is determined by the perturbation strength $\alpha$ according to $k = \alpha / 2n_0^2$.

The nanocavitiy was fabricated within a 250nm thick membrane of InGaAsP with six 75Å quantum wells (QW) positioned at the center. After the resonator pattern was etched into the active material, the original InP substrate was removed and the membrane was transferred to a sapphire plate using an ultraviolet curable optical adhesive to improve the vertical confinement of the electromagnetic field in the device. A detailed description of the fabrication process is given in Ref. 13. The effective index of the membrane was found to be approximately 2.8 for the $H_z$ polarization and 2.09 for the $E_z$ polarization. Since the $H_z$ polarization is more confined than the $E_z$ polarization and the optical gain of the compressively strained QW structure used favors the $H_z$ polarization[14], we optimized the radial structure to this polarization.

The emitted spectrum and the near-field (NF) intensity pattern of the nanocavity were examined at room temperature under pulsed optical pumping using a mode-locked Ti:Sapphire laser operating at 890nm. The pump beam was focused through the transparent sapphire substrate on the backside of the sample with a 50X objective lens. The position of this lens was used to control the size and the position of the pump spot. 50% of the pump beam intensity was split by a 3dB beam-splitter and was focused on a broad area detector to obtain the pump beam intensity. A 20X objective lens was used to collect the vertical emission from the front side of the sample and to couple the light into a multi-mode fiber to obtain the emitted spectrum or to focus it on an IR camera to obtain the NF intensity pattern.

Figure 2 depicts the emitted spectra from the $m=0$ laser for various pumping levels above the lasing threshold. The emitted light consists primarily of a single wavelength at $\lambda=1.56\mu m$, very close to the target design wavelength of $1.55\mu m$. The inset of Fig. 2 shows an L-L curve of the same device, indicating a threshold at $P_{th}=900\mu W$. It should be noted that the pump powers quoted indicate the overall power carried by the pump beam while the actual power absorbed by the QWs is significantly lower. At high pump levels (~1.5x$P_{th}$), two additional low-intensity emission lines appear at longer wavelengths (~1.59$\mu m$) although the main emission line remained the dominant one. We attribute these modes to emission from the external Bragg grating region[15].

Figure 3 shows a contour plot of the nanolaser index profile superimposed on a cross-section of the modal field intensity profile in the center of the active medium. As shown in the figure, the modal profile of the nanocavity is confined almost completely in the 300nm wide central pillar with a modal volume of $0.213(\lambda/n)^3$ (0.024 $\mu m^3$) – only 1.75 times the theoretically possible limit of a cubic half wavelength. This modal volume is a 30% lower than the modal volume demonstrated in PC defect cavities[16, 17]. This non-negligible improvement stems from the optimized match between the cavity dimensions and the (quasi) periodicity of the grating, made possible because of the cylindrically symmetric geometry.

The theoretical limit of modal volume can be achieved in principle using a spherical Bragg cavity[18] or a metallic box. However, in the geometry considered here (planar confinement by Bragg reflection and vertical confinement by TIR) there is a tradeoff between the vertical and horizontal confinements which limits the smallest possible $V_{mode}$. This tradeoff can be intuitively understood as follows: For a given membrane thickness, the transverse design and mode profile of the cavity are determined primarily by the effective index of the slab. The larger the effective index, the smaller the cavity, and correspondingly, the modal volume. On the other hand, larger effective indices require thicker membranes which, in turn, increase the modal volume. These two opposite processes create an optimal membrane thickness for which $V_{mode}$ is minimal.

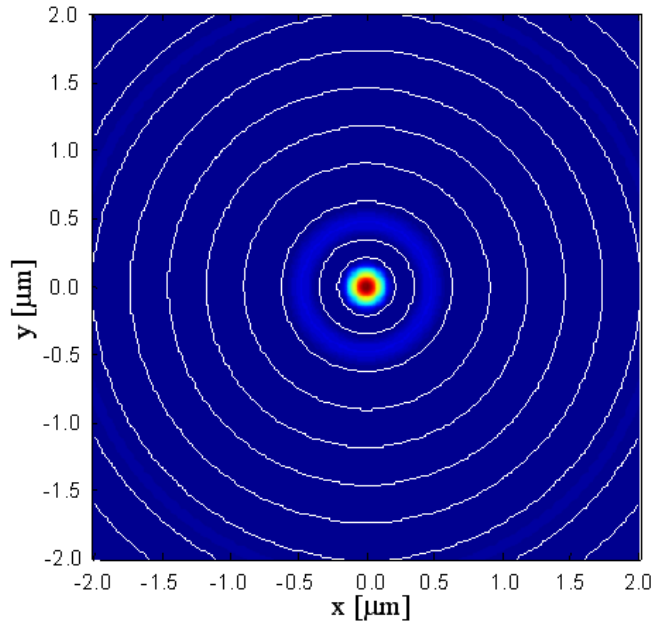

Fig. 3 (Color online) Calculated modal intensity profile of the nanocavity shown in Fig. 1.

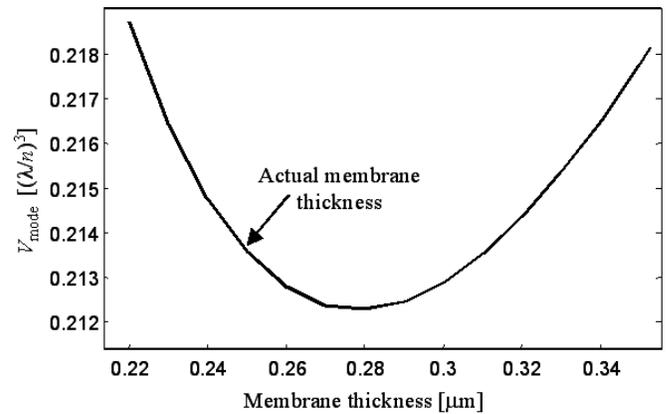

Fig. 4. The dependence of the normalized modal volume on the membrane thickness.

Figure 4 depicts the dependence of $V_{mode}$ on the thickness of the membrane. For the material system that was used, the smallest achievable modal volume is $0.212(\lambda/n)^3$ occurring for a membrane thickness of 0.28μm, which is less than one percent smaller than the $V_{mode}$ of the fabricated structure. The modal volume of the nanocavity could be further reduced to almost the theoretical limit if a first Bragg order scheme were used instead of the second order scheme (1).

In conclusion, we demonstrated single mode lasing at telecommunications wavelengths from a circular Bragg nanolaser with an ultra-small modal volume. Lasing was achieved at room temperature under pulsed optical pumping conditions at sub mW threshold levels. Such cavities can easily be integrated with other photonic devices such as PC waveguides and DFB lasers to realize compact and highly functional optical circuits.

REFERENCES

[1] C. K. Madsen and J. H. Zhao, Optical Filter Design and Analysis: A Signal Processing Approach. New York: Wiley, 1999.
[2] B. E. Little, Opt. Lett. 23, 1570 (1998).
[3] C. Y. Chao and L. J. Guo, Appl. Phys. Lett. 83, 1527 (2003).
[4] See for example K.J. Vahala, Nature 424, 839 (2003) and references therein.
[5] V. Van, T. A. Ibrahim, P. P. Absil, F. G. Johnson, R. Grover, and P. T. Ho, IEEE J. Sel. Top. Quantum Electron. 8, 705 (2002).
[6] Y. Akahane, T. Asano, B. S. Song and S. Noda, Nature 425, 944 (2003).
[7] M. Notomi, A. Shinya, S. Mitsugi, E. Kuramochi and H. Y. Ryu, Optics Express 12, 1551 (2004).
[8] M. Loncar, M. Hochberg and A. Scherer, Opt. Lett. 29, 721 (2004).
[9] J. Scheuer and A. Yariv, IEEE J. Quantum Electron. 39, 1555 (2003).
[10] J. Scheuer and A. Yariv, J. Opt. Soc. Am. B. 20, 2285 (2003).
[11] J. Scheuer, W. M. J. Green, G. DeRose and A. Yariv, Opt. Lett. 29, 2641 (2004).
[12] J. Scheuer and A. Yariv, Phys. Rev. E. 70, 036603 (2004).
[13] W. M. J. Green, J. Scheuer, G. DeRose and A. Yariv, J. Vac. Sci. Technol. B. 22, 3206 (2004).
[14] L. A. Coldren and S. W. Corzine, Diode Lasers and Photonic Integrated Circuits. New York: Wiley, 1995.
[15] J. Scheuer, W. M. J. Green, G. DeRose and A. Yariv, IEEE J. Sel. Top. Quantum Electron. 11, 476 (2005).
[16] O. Painter, R. K. Lee, A. Scherer, A. Yariv, J. D. O'Brien, P. D. Dapkus and I. Kim, Science 284, 1819 (1999).
[17] T. Yoshie, O. B. Shchekin, H. Chen, D. G. Deppe and A. Scherer, IEICE Trans. Electron. E87-C, 300 (2004).
[18] J. D. Jackson, Classical Electrodynamics 3rd ed. New York: Wiley, 1999.